%% file: PLUTO_v1_arxiv_May_2024.tex
\documentclass[10pt,twocolumn,letterpaper,leqno]{article}

\usepackage[pagenumbers]{cvpr} 
\usepackage{makecell}
\usepackage{caption}

\input{preamble}

\definecolor{cvprblue}{rgb}{0.21,0.49,0.74}
\def\thickhline{\noalign{\hrule height1.2pt}}
\usepackage[pagebackref,breaklinks,colorlinks,citecolor=cvprblue]{hyperref}
\usepackage{pgf}
\makeatletter
\robustify\@latex@warning@no@line
\makeatother
\usepackage{authblk}

\title{PLUTO: Pathology-Universal Transformer}

\author[1]{Dinkar Juyal}
\author[1]{Harshith Padigela}
\author[1]{Chintan Shah}
\author[2]{Daniel Shenker}
\author[2]{Natalia Harguindeguy}
\author[3]{Yi Liu}
\author[3]{Blake Martin}
\author[3]{Yibo Zhang}
\author[3]{Michael Nercessian}
\author[3]{Miles Markey}
\author[3]{Isaac Finberg}
\author[3]{Kelsey Luu}
\author[3]{Daniel Borders}
\author[3]{Syed Ashar Javed}
\author[3]{Emma Krause}
\author[ ]{Raymond Biju}
\author[ ]{Aashish Sood}
\author[ ]{Allen Ma}
\author[ ]{Jackson Nyman}
\author[ ]{John Shamshoian}
\author[ ]{Guillaume Chhor}
\author[ ]{Darpan Sanghavi}
\author[ ]{Marc Thibault}
\author[ ]{Limin Yu}
\author[ ]{Fedaa Najdawi}
\author[ ]{Jennifer A. Hipp}
\author[ ]{Darren Fahy}
\author[ ]{Benjamin Glass}
\author[ ]{Eric Walk}
\author[ ]{John Abel}
\author[ ]{Harsha Pokkalla}
\author[ ]{Andrew H. Beck}
\author[4]{Sean Grullon\thanks{Corresponding Author: sean.grullon@pathai.com}}
\affil[1]{Project Lead and Co-lead author}
\affil[2]{Workstream Lead}
\affil[3]{Core Team}
\affil[4]{Directional Lead}
\affil[ ]{PathAI}

\begin{document}
\maketitle
\let\thefootnote\relax\footnote{The authors would like to thank Jon Ross, Elliot Miller, Alfred Eng and Maximillian King for their MLOps support, Jacqueline Brosnan-Cashman for her help with illustrations and thank Vincent Billaut and Michael Griffin for their manuscript feedback}

\begin{abstract}
Pathology is the study of microscopic inspection of tissue, and a pathology diagnosis is often the medical gold standard to diagnose disease. 
Pathology images provide a unique challenge for computer-vision-based analysis: a single pathology Whole Slide Image (WSI) is gigapixel-sized and often contains hundreds of thousands to millions of objects of interest across multiple resolutions. 

In this work, we propose PathoLogy Universal TransfOrmer (PLUTO): a light-weight pathology FM that is pre-trained on a diverse dataset of 195 million image tiles collected from multiple sites and extracts meaningful representations across multiple WSI scales that enable a large variety of downstream pathology tasks. In particular, we design task-specific adaptation heads that utilize PLUTO's output embeddings for tasks that span pathology scales ranging from subcellular to slide-scale, including instance segmentation, tile classification, and slide-level prediction. 
We compare PLUTO’s performance to other state-of-the-art methods on a diverse set of external and internal benchmarks covering multiple biologically relevant tasks, tissue types, resolutions, stains, and scanners. 
We find that PLUTO matches or outperforms existing task-specific baselines and pathology-specific foundation models, some of which use orders-of-magnitude larger datasets and model sizes when compared to PLUTO. 
Our findings present a path towards a universal embedding to power pathology image analysis, and motivate further exploration around pathology foundation models in terms of data diversity, architectural improvements, sample efficiency, and practical deployability in real-world applications.
\end{abstract}  

\section{Introduction}
\label{sec:intro}

Pathology as a medical discipline is instrumental in providing diagnostic and prognostic information to clinicians and patients. 
In a pathology workflow, surgical tissue specimens are collected, stained, and fixed for microscopy. 
Microscopic analysis of the tissue is used to establish a diagnosis, estimate disease severity, and identify relevant clinical features for treatment \citep{walk2009role, MADABHUSHI2016170, bejnordi2017lymph}.

The practice of pathology is not inherently digital; traditionally, pathology slides are manually examined under a microscope. 
Microscopy slides are increasingly being digitized in their entirety via slide scanning, generating digital whole slide images (WSIs or slides).  
While WSIs provide a wealth of information about a specimen to trained readers such as pathologists, the images themselves are enormous. 
Each WSI contains up to millions of cells and can be gigapixels in scale, making an exhaustive quantitative manual analysis of WSIs nearly impossible. 
In addition, information for making pathologic decisions or classifications may exist at multiple scales, from several $\mu$m to several cm, complicating analysis.

Artificial intelligence (AI) and machine learning (ML) techniques are well-suited for the quantitative study of these extremely large WSIs.
A wide variety of ML techniques have been developed and applied to the pathology domain, ranging from detection and characterization of microscopic biological entities within the WSI, to end-to-end frameworks for making slide-level predictions or diagnoses \citep{wang2016deep, bulten2020automated, campanella2019clinical, diao2021human, bosch2021machine, javed2022additive, chen2022scaling}.

However, developing supervised machine learning models for pathology comes with many challenges \cite{tizhoosh2018artificial}. These algorithms need large amounts of labeled data which is often expensive to collect and in some cases hard to source due to the low prevalence of disease characteristics. Additionally, these models need to generalize across variations introduced by source sites, scanners, and staining procedures \cite{Tellez_2019, s_dota, nguyen2024contrimix, li2022selftraining}. Lowering the data burden and improving the robustness of these models is important for broad-scale adoption of AI models in pathology practice. Furthermore, the diversity of individual tasks in pathology (such as classification, segmentation, and slide-level prediction) makes training bespoke AI models from scratch challenging. 

Foundation models \cite{fms} (FMs) are large scale deep-learning models often pre-trained on broad-scale, unlabeled data using self-supervision and can be adapted to multiple downstream tasks \cite{gpt4, dinov2} such as image classification and object detection with fewer labels than traditional strongly-supervised methods \cite{fm-fewshot}. This adaptation procedure normally involves utilizing the representation (also referred to as embeddings) produced by a pre-trained FM \textit{backbone} to fine-tune a task head (with significantly fewer model parameters than the backbone) on a particular downstream task.

\subsection{Previous Pathology Foundation Models}

\begin{figure*}
    \centering
    \includegraphics[width=\textwidth]
    {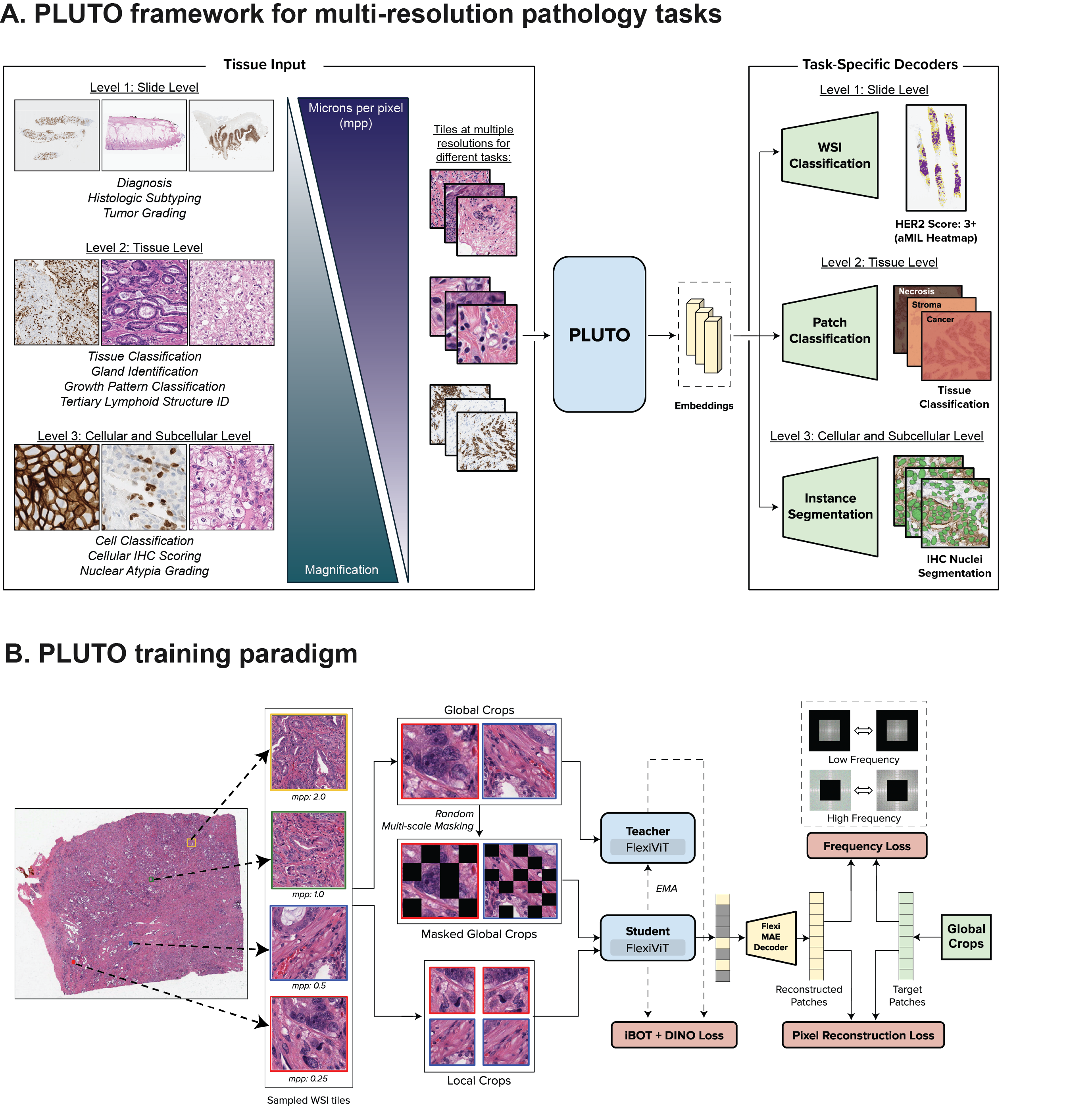 }
    \caption{Overview of PLUTO. Panel A) outlines the PLUTO multi-resolution adaptation pipeline. Tiles are extracted from WSIs at multiple resolutions and correspond to scales that capture  different biological contexts. We organize pathology tasks according to these biological contexts as slide level, tissue level, and cellular \& subcellular level tasks, respectively. PLUTO generates embeddings that are task-agnostic and can be used in a variety of downstream tasks, where adaptation to WSI-level prediction, tile classification, and instance segmentation are shown. Panel B) demonstrates the PLUTO architecture in detail. WSI tiles at multiple resolutions are masked with varying patch sizes and passed to the backbone for self-supervised pre-training. The architecture is optimized for flexibility across multiple scales and patch sizes. In addition to DINO and iBOT losses, MAE and Fourier losses are applied across varying mask sizes to control the amount of low- and high-frequency information that is preserved.}
    \label{fig:FM_arch}
\end{figure*}

FMs are promising for pathology as they can take advantage of large amounts of unlabeled data to build rich representations which can be easily adapted for downstream tasks in a data-efficient manner. The diversity of pre-training data results in these models generating robust representations, enabling them to generalize better than individual task-specific models trained on smaller datasets \cite{fms}.  

Additionally, sharing a backbone across different tasks could also reduce the development and maintenance overhead associated with bespoke task-specific models. Given this prospect, the computational pathology community has made rapid progress in applying self-supervised techniques  that have shown promise on natural images such as DINO \cite{dino}, iBOT \cite{ibot}, and DINOv2 \cite{dinov2} to pathology. Most of these efforts have relied on pre-training with a large amount of proprietary data and scaling up the number of backbone parameters used in order to demonstrate high performance on various downstream tasks including tissue classification, disease subtyping classification, and cancer histology segmentation. 

 Kang \textit{et al.} \cite{lunit} have compared a panel of self-supervised learning strategies on tile classification of tissue types and cell segmentation, and have reported that the backbone trained with the DINO \cite{dino} vision transformer backbone demonstrated superior performance. Filiot \textit{et al.} have developed Phikon \cite{phikon} based on the iBOT \cite{ibot} self-supervised strategy, relying on public data from The Cancer Genome Atlas (TCGA) \cite{tcga}, and evaluated their model on tissue type tile classification and weakly-supervised slide-level cancer subtyping. Shaikovski \textit{et al.} have developed Virchow \cite{virchow} based upon the DINOv2 vision transformer (ViT) \cite{dosovitskiy2021image} backbone, and trained the model at scale with two billion tiles corresponding to $1.5$ million WSIs from the Memorial Sloan Kettering Cancer Center (MSKCC). This model has been evaluated on a tile classification task across $17$ different cancer types including seven rare cancer types, and has also demonstrated high performance on slide-level biomarker prediction tasks. Winterhoff \textit{et al.} have developed RudolfV \cite{rudolfv}, also based on DINOv2 pre-training and relied on both public data from TCGA and proprietary data. Winterhoff \textit{et al.} have reported tissue tile classification results across six public datasets and cell segmentation benchmarks on their proprietary test set. UNI \cite{uni} from Chen \textit{et al.} also used a DINOv2 backbone, but introduced additional data diversity through the ``Mass-100K'' dataset which was sourced from Massachusetts General Hospital (MGH), Brigham and Women’s Hospital, and the Genotype–Tissue Expression (GTEx) consortium. UNI has been adapted to a wider variety of tasks in addition to tissue classification, including disease subtyping classification and cancer histology segmentation.

Despite the encouraging results from these pathology FMs, there are several areas of improvement that would further drive the adoption of FMs in pathology practice. First, FM pre-training has predominantly relied on a large amount of proprietary data from a single site. There are site-specific batch effects \cite{pathreview} in WSIs arising from site-specific variation, including both stain and the patient population, which can lower the robustness of AI models when unaccounted for. Additionally, a lack of pixel-level annotations at scale can limit the downstream performance of these pathology FMs as recent studies \cite{ssldiversity} have demonstrated the importance of data and label diversity in self-supervised pre-training. Our large network of pathologists has gathered millions of pixel level annotations \cite{oghif}, which provided us with a powerful source of data to drive FM improvements. Second, the architectural design of pathology FMs could further take advantage of the multi-scale nature of WSIs outlined in Section \ref{sec:pyramid}. Finally, FM backbones have often been trained with a large number of model parameters, which increases the complexity and cost of deploying these models and hence limits their practical use in routine pathology practice.

\subsection{Our Approach}
\label{sec:approach}

We designed and built the \textit{PathoLogy-Universal TransfOrmer}, or PLUTO, a state-of-the-art pathology foundation model that, inspired by the dwarf planet, is based on a novel light-weight ViT backbone that is pre-trained on a diverse dataset from multiple sites and extracts meaningful representations across the levels of the WSI pyramid outlined in Section \ref{sec:pyramid}. The key features of PLUTO, namely the pre-training dataset, architecture, multi-scale task evaluation, and deployability, are outlined below:

\begin{enumerate}
    \item \textbf{Pre-training Dataset} We compiled a large dataset across a diverse spectrum of histology stains, scanners, and biological objects across resolution scales that include $200$+ biologically-meaningful objects and region types (which we term \textit{substances}) from more than $50$ sources (Section \ref{sec:data}).
    \item \textbf{Architecture} We designed the PLUTO backbone to generate informative feature representations at different length scales from a compact ViT backbone. We achieved this by implementing a self-supervised learning scheme that accommodates flexible patch sizes from the FlexiViT scheme \cite{flexivit}, extending it to accommodate multiple magnifications during training, and modifying the DINOv2 loss by adding a Masked Autoencoder (MAE) \cite{mae} objective and a Fourier-loss-based term to modulate the preservation of low- and high-frequency components (Section \ref{sec:backbone}).
    \item \textbf{Multi-scale Evaluation} We evaluated the quality of the resulting FM by constructing a suite of adaptation heads to perform diverse, challenging tasks across the levels of the WSI pyramid (Sections \ref{sec:pyramid}, \ref{sec:method_adapt}), and evaluated performance across different biologically-relevant benchmarks (Section \ref{sec:res}).
    \item \textbf{Deployability.} Performing a computational pathology task may require embedding tens to hundreds of thousands of WSI tiles to make a single prediction. To enable this, we focused on developing a model that was efficient (Section \ref{sec:deploy}).
\end{enumerate}

\section{Pathology Task Hierarchy and Pre-training Dataset}
\subsection{Pyramid Structure of Whole Slide Images}
\label{sec:pyramid}

\begin{figure*}
    \centering    
    \includegraphics[width=0.9\textwidth]{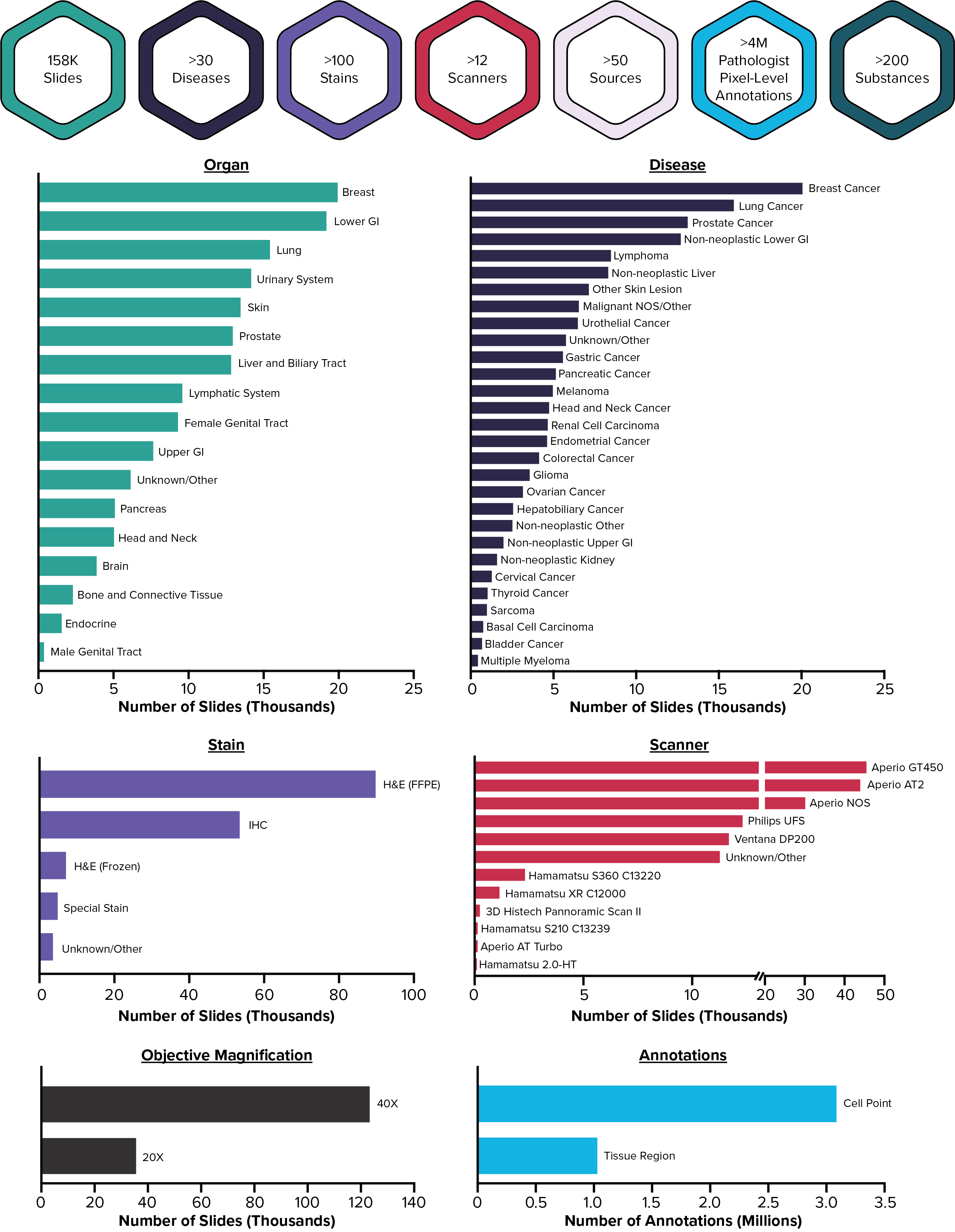}    

    \caption{Dataset characterization for the pre-training dataset. The distribution of the dataset by organ, disease, stain, scanner, and objective magnification is shown, as well as the distribution of cell point and tissue region annotations which augment the pre-training dataset (NOS: Not Otherwise Specified). Aggregate data characteristics are summarized above these distributions which also indicate the number of biologically-meaningful objects and region types, which we term \textit{substances} (e.g. lymphocyte, blood vessel, Gleason pattern 3 prostate cancer, tumor bed). The large number of source sites (50+) guarantees large diversity during PLUTO self-supervised pre-training.}
    \label{fig:main_metadata_fig}
\end{figure*}

WSIs are digitized and stored in a multi-scale pyramidal structure, where the base of the pyramid is the highest-resolution image data as captured by the slide scanner. The resulting scan of a WSI can reach $200,000\times200,000$ pixels at a full resolution of $0.25$ microns ($\mu$m) per pixel (mpp) \cite{sellaro}; however, different ``levels'' of the pyramid may be accessed for different purposes.

Biological entities observed on WSIs vary dramatically in scale, and therefore pathologists will commonly move between magnifications to assess different aspects of a tissue sample on a pathology slide \cite{molin}.
At low magnification, pathologists may scan across slides to identify regions of interest in the tissue, with characteristic lengths of approximately $1$ mm–$1$ cm.
At middle magniﬁcation (such as $5$–$10\times$) pathologists commonly view structures at length scales of $200$ $\mu$m–$1$ mm. At this scale, pathologists distinguish between tissue types, glands, tumor growth patterns, histologic subtypes of diseases, or other multicellular entities in the image.
At high magnification ($20$–$40\times$) it is possible to resolve entities $1$ $\mu$m–$50$ $\mu$m in length, such as individual cell identities, subcellular structural morphology used in determining malignancy, or localization of immunohistochemical (IHC) staining \cite{ihc}.

The hierarchical nature of biological entities necessitates considering the multiple scales at which information must be extracted and used by ML algorithms. For example, passing a $224\times224$ image tile at $0.25$ mpp through an encoder developed for encoding at $1$ mpp may completely miss relevant nuclear pleomorphism, whereas passing a $224\times224$ tile at $1$ mpp through an encoder developed for encoding at $0.25$ mpp may be unable to adequately distinguish between acinar and lepidic growth patterns. For clarity, we organize pathology tasks according to such biological scales as follows:
\begin{itemize}
\item \textbf{Level 1: Slide Level} This scale includes tasks that label the entire slide such as predicting driver gene mutations in cancer, histologic subtyping, or tumor grading. However, it is uncommon that slide-level assessments are made at slide-level magnification. Typically, assessments made at this scale are aggregated across evaluation of higher-magnification tiles.
\item \textbf{Level 2: Tissue Level} This is the scale at which it is possible to identify and characterize tissue regions (e.g. cancer regions and necrotic regions) and many-cellular objects such as glands.
\item \textbf{Level 3: Cellular and Subcellular Level} This is typically the maximal resolution of a WSI, where cellular and subcellular morphology is evident. 
\end{itemize}

\subsection{Pre-training Data Characteristics}
\label{sec:data}

The dataset used for self-supervised pre-training comprises public and proprietary datasets, totaling $195$M image tiles sampled at four resolutions from $158,852$ WSIs derived from over $50$ source sites (Figure \ref{fig:main_metadata_fig}). The WSIs span over $16$ tissue groups (Table \ref{table:organ}) and $28$ disease areas (Table \ref{table:disease}), which capture a broad range of benign, malignant, and inflammatory lesions. Additionally, the training set is unique in the representation covering $11$ scanners and four stain groups: hematoxylin and eosin (H\&E) formalin-fixed paraffin-embedded (FFPE), H\&E frozen, IHC (capturing over $100$ distinct IHC stains including PD-L1, HER2, and Ki-67) and special stains (including six stains such as trichrome and iron) (Tables \ref{table:scanner}, \ref{table:stain}). The base objective magnification of our training set consists of both $20\times$ and $40\times$ slides (Table \ref{table:objective}). 

Prior to tile sampling, we identify usable tissue regions using ArtifactDetect \cite{artifactdetect}, a ResNet-style convolutional neural network trained to segment usable tissue, artifact, and background. This model provides significant performance gains for usable tissue identification in our experience compared to other methods such as Otsu masking in terms of generalization across different stains, scanners, diseases, and organs. Tiles are sampled from regions of usable tissue at up to four different resolutions: $40\times$ ($0.25$ mpp), $20\times$ ($0.5$ mpp), $10\times$ ($1$ mpp), and $5\times$ ($2$ mpp). 

Following findings from DINOv2 \cite{dinov2} highlighting the significant value of incorporating curated data into self-supervised pre-training, this dataset is augmented with an additional set of samples extracted from over four million manual annotations from board-certified pathologists. These hand-drawn pathologist annotations correspond to hundreds of different types of biological entities at various scales (e.g., lymphocyte, blood vessel, Gleason pattern 3 prostate cancer, tumor bed). During pre-training, the labels are discarded, but the inclusion of pathologist-curated regions covering a wide range of biological patterns provides an implicit data diversity in the pre-training process. This source of biological diversity, combined with the broad range of stains, organs, diseases, and source sites, makes this one of the most diverse large-scale digital pathology datasets to date.

\section{Methods}
\subsection{PLUTO Architecture Overview}
\label{sec:backbone}

PLUTO (Figure \ref{fig:FM_arch}) is designed with specific characteristics in mind to enable its usage on a wide range of use-cases, as described in Section \ref{sec:approach}.
    To design the PLUTO architecture, we start from DINOv2 \cite{dinov2} which combines DINO \cite{dino} and iBOT \cite{ibot} losses (along with KoLeo regularizer) to learn relevant representations at the tile and patch levels respectively. Note, we use \textbf{tile} and \textbf{image} interchangeably to refer to the image tiles and \textbf{patches} to refer to the patch-tokens obtained by dividing the tile into smaller patches for processing in ViTs. The DINOv2 architecture was primarily developed for natural images that are often object-centric. Pathology WSI tiles on the other hand  have thousands of objects such as nuclei, cells, and glands with different sizes, observed at different image resolutions (Section \ref{sec:pyramid}).   
To design an encoder which can capture details of objects at different levels of granularity, we add in a MAE \cite{mae} objective with multi-scale masking. The MAE setup tries to reconstruct masked regions of the input image (often a large fraction of the input) from the unmasked regions. We perform masking by varying the patch sizes used for masking while using images across different resolutions of the WSI as shown in Figure \ref{fig:FM_arch}. In addition to the pixel-level reconstruction loss used in MAE, we add a Fourier reconstruction loss to control the amount of low- and high-frequency information preserved during the pre-training process.

To enable the encoder and decoder to handle varying patch sizes for multi-scale masking, we employ the FlexiViT setup \cite{flexivit}. Since patch size controls the granularity of information captured by the encoder, different downstream tasks may need different patch sizes for optimal performance. The FlexiViT setup allows us to adapt the same backbone to different tasks without needing to train a backbone for every patch size. The patch size also determines the effective sequence length used in ViTs and FlexiViT allows us to cater to different compute budgets by selecting the most suitable patch size at inference time. 

\subsection{PLUTO Backbone Training}

We extract images from the dataset described in Section \ref{sec:data}, with the image resolution selected randomly (with pre-specified probabilities). Two global crops and four local crops of sizes $224$ and $96$ respectively are taken from each image, consistent with DINOv2 training. The local crops are passed to the student, while the global crops are passed to the teacher. The teacher's weights are updated using an exponential moving average of the student's weights rather than backpropagation. The crops provided to the student are randomly masked for the iBoT objective. Meanwhile, a separate masking setup with a higher masking ratio is applied to the global crops for the MAE objective. The mask sizes are consistent with the patch sizes which are dynamically chosen from [$8$, $16$, $32$] to enable multi-scale masking. The flexible patch embedding step ensures that the architecture can accommodate patches of variable sizes. Since a vanilla MAE decoder cannot work with variable sized masks due to the presence of linear layers, a similar ``flexification'' setup is added to the MAE decoder to generate reconstructions with variable mask sizes. Learnable position embeddings are used in the MAE decoder. L2-norm loss between the reconstructed and the original image is used for the MAE objective. 

Beyond the original MAE objective, we decompose the reconstructed image into its low- and high-frequency components. This decomposition is crucial for addressing distinct aspects of image quality that are captured in different frequency ranges. To achieve this, the Fourier spectrum of the reconstructed image is dissected into low- and high-frequency bands using a set of low-pass and high-pass filters. By applying these filters, the method effectively isolates the components of the image that represent basic structures and details (low frequency) from those encapsulating finer details and textures (high frequency). After this separation, the L2 loss is computed independently for both the low- and high-frequency parts of the image. This bifurcated approach allows for a more nuanced adjustment and optimization of the reconstructed image by applying tunable weights to the losses from each frequency band before their aggregation. The sum of these weighted losses forms the overall Fourier reconstruction loss, which the training process aims to minimize \cite{FreMIM}. The whole loss function $\mathcal{L}(\hat{y}, y)$ is listed in Eq.~\eqref{eq:1} and the detailed Fourier loss $\mathcal{L}_{\text{Fourier}}(\hat{y}, y)$ is listed in Eq.~\eqref{eq:2}. $\hat{y}$ represents the masked regions of the predicted image and $y$ represents the masked regions of the ground truth image. The Discrete Fourier Transform (DFT) is denoted by $\mathcal{F}$. The mask $M$ in the Fourier space acts as a low-pass filter, and $1-M$ acts as a high-pass filter. The weights $\lambda_1$ and $\lambda_2$ are used to balance the contributions of the low-pass and high-pass filtered errors, respectively. The value of $\lambda_1$ is set to 5 and $\lambda_2$ is set to 1. The weights of other losses are set to 1. $\|\cdot\|_2$ denotes the L2 loss.

\begin{equation} 
    \begin{split}
    \mathcal{L}(\hat{y}, y) &= \mathcal{L}_{\text{DINO}}(\hat{y}, y) + \mathcal{L}_{\text{iBOT}}(\hat{y}, y) \\
    &\quad + \mathcal{L}_{\text{MAE}}(\hat{y}, y) + \mathcal{L}_{\text{Fourier}}(\hat{y}, y)
    \end{split} \label{eq:1}
\end{equation}

\vspace{-1.8\baselineskip} 

\begin{multline} 
    \mathcal{L}_{\text{Fourier}}(\hat{y}, y) = \lambda_1 \cdot \| M \cdot \mathcal{F}(\hat{y}) - M \cdot \mathcal{F}(y) \|_2^2 \\
    + \lambda_2 \cdot \| (1-M) \cdot \mathcal{F}(\hat{y}) - (1-M) \cdot \mathcal{F}(y) \|_2^2 \label{eq:2}
\end{multline}
 
We observe slightly better performance with the teacher over the student, and thus use the teacher for all downstream tasks.

We use ViT-S for the student and teacher encoders, and a shallower model is used for the MAE decoder. For training, we use AdamW with a base learning rate of $0.002$ and a learning rate warmup for the first 5 epochs. We use a distributed training setup to scale the training across $64$ NVIDIA A40 GPUs.

\subsection{PLUTO Adaptation}
\label{sec:method_adapt}
The backbone training process outlined above learns generic, task-
features. In order to leverage its general capabilities, we add task-specific heads and \textit{adapt} these heads through supervised fine-tuning, while keeping the backbone fixed, or frozen. This adaptation process is efficient and provides the flexibility to use the same pre-trained backbone for specialized tasks across the biological scales described in Section \ref{sec:pyramid}. Although different tasks may require the use of different patch sizes to capture relevant context, the FlexiVit setup allows us to dynamically select the backbone patch size for adaptation.

In this section we describe the techniques underlying the specialized task heads that work across these biological scales. At Level 1, we adapt PLUTO to slide-level classification tasks with a multiple-instance learning (MIL) head, further discussed in Section \ref{sec:mil}. We adapt PLUTO to tissue-level (Level 2)  \& cellular- and subcellular-level (Level 3) biological scales, respectively, through the tile-level classification and instance segmentation task heads that we further describe in Section \ref{sec:segmentation_adaptation}. The overall architecture is illustrated in panel B of Figure \ref{fig:FM_arch}.

\subsubsection{Slide-level Task Adaptation}
\label{sec:mil} We adapt PLUTO to Level 1 slide-level tasks described in Section \ref{sec:pyramid} by performing weak supervision on slide-level labels. In particular, MIL \cite{ilse2018attention} is a weakly supervised learning technique where sets of instances are grouped into a ``bag'' and used to learn bag-level labels. These MIL models consist of three parts: (1) a featurizer which generates representations of each image tile in a bag, (2) an aggregation module which combines tile representations using a permutation-invariant function (typically attention) to generate a bag-level representation, and (3) a classifier which outputs a bag-level prediction. We adapt our FM backbones by using the pre-trained backbones directly as featurizers, with the adaptation heads (which are the downstream components consisting of an attention module and classifier layer) operating on the feature vectors generated from these backbones. These models are trained with the featurizer either frozen (FZ) or unfrozen (FT, or fine-tuned) during MIL training. We use the AdditiveMIL classifier \cite{javed2022additive}, which enables interpretable model predictions and class-wise heatmaps.

\subsubsection{Tissue-level, Cellular- and  Subcellular-level Task Adaptation}
\label{sec:segmentation_adaptation}

\begin{figure*}
    \includegraphics[width=\linewidth]{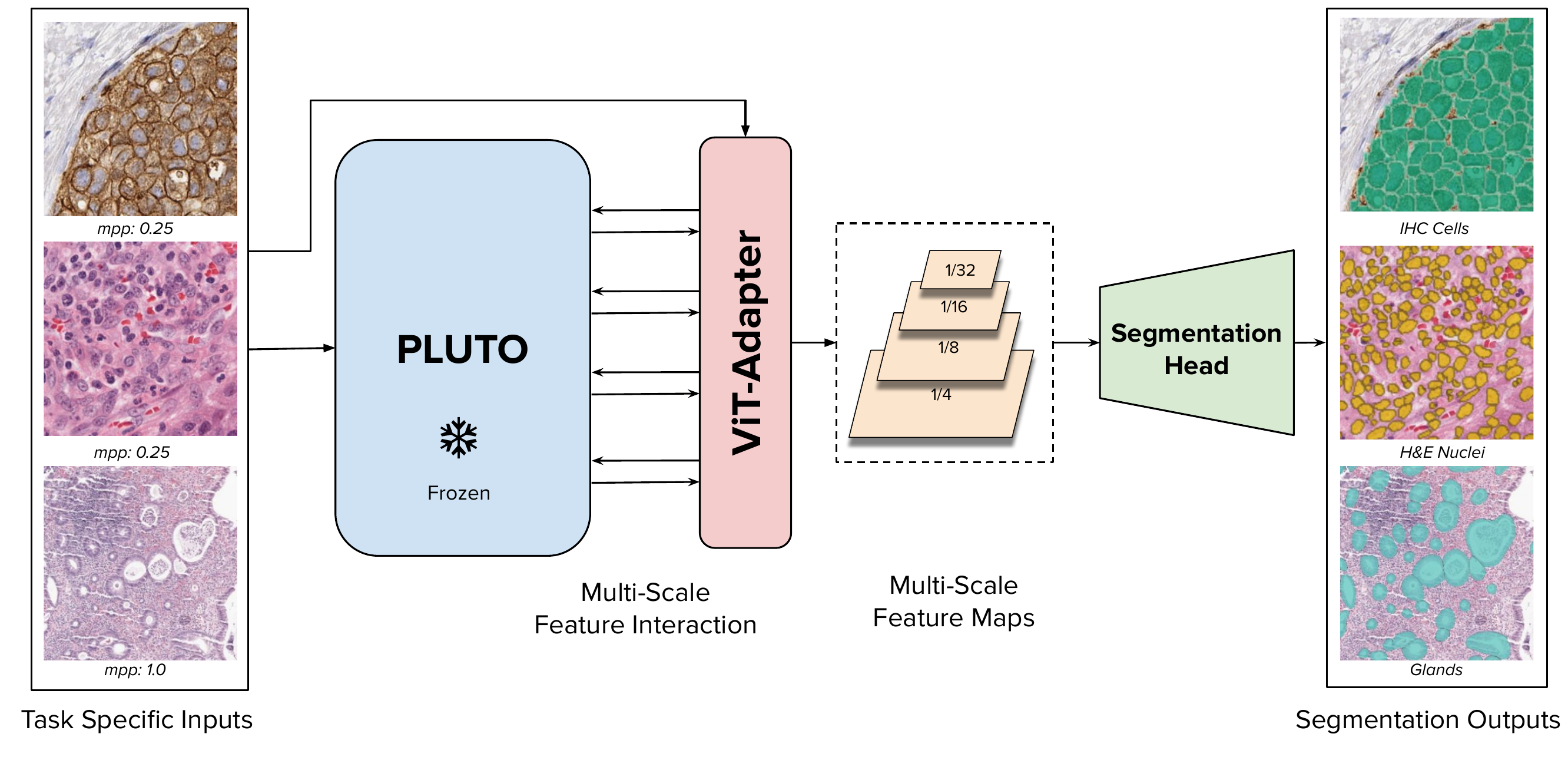}
    \caption{Instance segmentation adaptation with PLUTO. In this figure, we demonstrate example task-specific inputs and outputs using the frozen PLUTO backbone, and our segmentation adaptation approach on top using an adapter that outputs maps at varying spatial and semantic resolutions, followed by a segmentation head to generate instance segmentation masks. We demonstrate that our approach works across object scales from nuclei (top two images) to glands (bottom) and across stain types. In our proprietary datasets, Gland segmentation is trained at $768\times768$ at $1$ mpp whereas nuclei segmentation is at $384\times384$ at $0.25$ mpp. In our public benchmark experiments, we follow the prescribed task setup.}
    \label{fig:adaptation_instance_seg}
\end{figure*}

We adapt PLUTO to tissue-level and cellular/subcellular-level biological scales through fine-tuning either a tile classification or an instance segmentation adaptation head. These two adaptation strategies are informed by the availability of labeled data: tile-level classification only requires labels at the image tile level, whereas instance segmentation requires pixel-level annotations. 

We explore and benchmark a range of adaptation head architectures for tile classification, ranging from single linear layers to multilayer perceptrons (MLPs) with different pooling strategies.  By incorporating domain-specific knowledge from histopathology tiles, the aim is to develop a light-weight adaptation head that identifies critical features captured in the backbone embeddings of these tiles, such as cellular morphology and tissue region characteristics. This allows for the classification of tiles into various categories, such as healthy tissue or cancerous tissue regions. 

In the analysis of gigapixel histopathology slides, such tile-level classification tasks can be effective substitutes for segmentation, a task which presents significant challenges in terms of collecting exhaustive, high-quality annotations. However, morphological descriptors of nuclei, cells, glands, vessels, and other biological entities are crucial prognostic indicators for various pathological analyses. Therefore, we developed specialized task heads designed for instance segmentation, enhancing our ability to effectively analyze and interpret these critical features.

We adapt the SSL-pre-trained ViT backbone to instance segmentation tasks via two distinct frameworks: Mask R-CNN \cite{maskrcnn} and Mask2Former \cite{mask2former}, as illustrated in Figure \ref{fig:adaptation_instance_seg}. While Mask R-CNN relies on region proposals and conventional mechanisms such as non-maximum suppression, Mask2Former employs a transformer-based approach, leveraging object queries to generate instance segmentation results, a methodology that presents a compelling advantage for our purposes since it does not rely on hand-tuned region proposals. To the best of our knowledge, this is the first work comparing Mask2Former to ViT + Mask R-CNN approaches on histopathology tasks.  We also experimented with combining the ViT with a ViT-Adapter \cite{vitadapter}, which has been shown to improve segmentation performance.  The output feature maps of the adapter, corresponding to different spatial resolutions of the input image, are used as the input to Mask R-CNN and Mask2Former. 

We report the performance of these adaptation approaches on both proprietary datasets and public datasets, as we present in Section \ref{sec:res}. We evaluate our adaptation techniques on our proprietary datasets with the object detection F1, semantic segmentation F1 and Aggregated Jaccard Index (AJI) \cite{7872382} metrics. We evaluate our methods on public datasets with the metrics as originally defined in those public benchmarks.

\section{Results}
\label{sec:res}

We evaluate the effectiveness of adapting PLUTO for tasks across biological scales corresponding to the pyramid levels of a WSI outlined in Section \ref{sec:pyramid}. A summary list of all benchmarks is given in Table \ref{table:all_benchmarks_overview}.

\begin{table*}
\centering
\resizebox{\textwidth}{!}{
\begin{tabular}{c c c c c c c c }
 \thickhline
 Task Level & Source & Benchmark Name & Task Type & \# Samples & \# Classes & Resolution & Patch Size\\
 \hline
 \hline
 Slide & Public & NSCLC subtyping \cite{tcga} & WSI-level Classification & $500$ & $2$ & $0.5$ mpp & $16$ \\
 Tissue & Public & H\&E CRC-100K \cite{NCTCRC} & Tile Classification  & $107,180$ & $9$ & $0.5$ mpp & $16$ \\
Tissue & Public & Camelyon17-WILDS \cite{camelyon17_wilds} & Tile Classification  & $455,954$ & $2$ & $1$ mpp & $16$ \\
Tissue & Public & GlaS \cite{glas} & Instance Segmentation  & $165$ & $2$ & $0.465$ mpp & $16$ \\
Cell & Public & PanNuke \cite{pannuke} & Instance Segmentation  & $481$ & $5$ & $0.25$ mpp & $16$ \\
 \hline
 Slide & Proprietary & HER2 Scoring & WSI-level Classification & $500$ & $4$ & $0.5$ mpp & $16$  \\
 Tissue & Proprietary & H\&E Pan-Onc. Tissue & Tile Classification & $196,807$ & $4$ & $0.5$ mpp & $16$ \\
 Tissue & Proprietary & IHC Pan-Onc. Tissue & Tile Classification & $196,807$ & $4$ & $0.5$ mpp & $16$ \\
 Tissue & Proprietary & H\&E IBD Tissue & Tile Classification & $648,794$ & $10$ & $0.5$ mpp & $16$ \\
Tissue & Proprietary & H\&E Glands & Instance Segmentation  & $518$ & $2$ & $1$ mpp & $8$, $16$, $32$  \\
Cell & Proprietary & H\&E Pan-Onc. Cell & Tile Classification & $212,036$ & $9$ & $0.5$ mpp & $8$ \\
Cell & Proprietary & IHC Cell & Instance Segmentation  & $225$ & $2$ & $0.25$ mpp & $8$, $16$  \\
Cell & Proprietary & IHC and H\&E Nuclei & Instance Segmentation  & $216$ & $2$ & $0.25$ mpp & $8$, $16$ \\
\thickhline
\end{tabular}}
\caption{Summary overview of PLUTO benchmarks. The benchmarks cover the range of pyramid levels in WSIs, as well as resolutions and task types. The variety of tasks demonstrates the adaptability of the pre-trained PLUTO embeddings. Patch Size denotes the FlexiViT patch size used for that downstream task.}
\label{table:all_benchmarks_overview}
\end{table*}

\subsection{Slide-level Results}
\subsubsection{Slide-level Datasets}
We consider two slide-level prediction tasks for evaluating our backbone. The first is the prediction of the cancer subtypes Adenocarcinoma and Squamous cell carcinoma in non-small cell lung carcinoma (NSCLC) H\&E-stained WSIs, a popular benchmark for slide-level evaluation. The second is the quantification of the HER2 biomarker across four scores ($0$, $1+$, $2+$, $3+$) in IHC-stained breast cancer slides, which measures the expression level of the HER2 protein and is clinically relevant for targeted patient therapy \cite{venetis2022her2}. 

For NSCLC subtyping, we use slides from the publicly available TCGA Adenocarcinoma (LUAD) and Squamous Cell Carcinoma (LUSC) groups. We use $500$ slides for model development and $247$ ($128$ LUAD / $119$ LUSC) slides for test set evaluation. We evaluate out-of-distribution (OOD) performance using a proprietary dataset of $205$ WSIs ($162$ Adenocarcinoma WSIs, $45$ Squamous Cell Carcinoma WSIs) collected from a different source site with varying image acquisition and processing steps, resulting in visual differences from the TCGA WSIs. Since slide-level prediction tasks are often limited by the number of slides available for development, we limit our development set to $500$ slides for both of these tasks and evaluate model performance on in-distribution (ID) and OOD test sets.

For HER2 scoring we use slides from an internal dataset constructed from multiple source sites, scanners and stain clones. We use $500$ slides for model development and $250$ slides with similar sample characteristics for ID evaluation. For OOD evaluation we use $229$ slides collected from two held-out source sites with different sample characteristics.

\subsubsection{Slide-level Results}

We compare our PLUTO pre-trained backbones against both pathology and ImageNet pre-trained baselines. We compare against an ImageNet pre-trained CNN-based backbone (ShuffleNet) \cite{ma2018shufflenet}, as well as Meta-DINOv2 ViT-S pre-trained on ImageNet \cite{dinov2}. During MIL training, the ViT-S based models are trained with frozen featurizers (FZ), and the PLUTO backbone is evaluated with a FlexiViT patch size of $16$. CNN-based models are trained with both frozen (FZ) and fine-tuned (FT) featurizers due to the smaller backbone size. 

We use macro-F1 and AUROC for evaluating the ID and OOD performance for both datasets. The metrics are computed using $1,000$ bootstrapped runs, reporting the mean and standard deviation. 
Results of these experiments are reported in Table \ref{table:MIL}. The pathology pre-trained backbones are the best performing on the ID test sets.

\begin{figure*}
    \includegraphics[width=\linewidth]{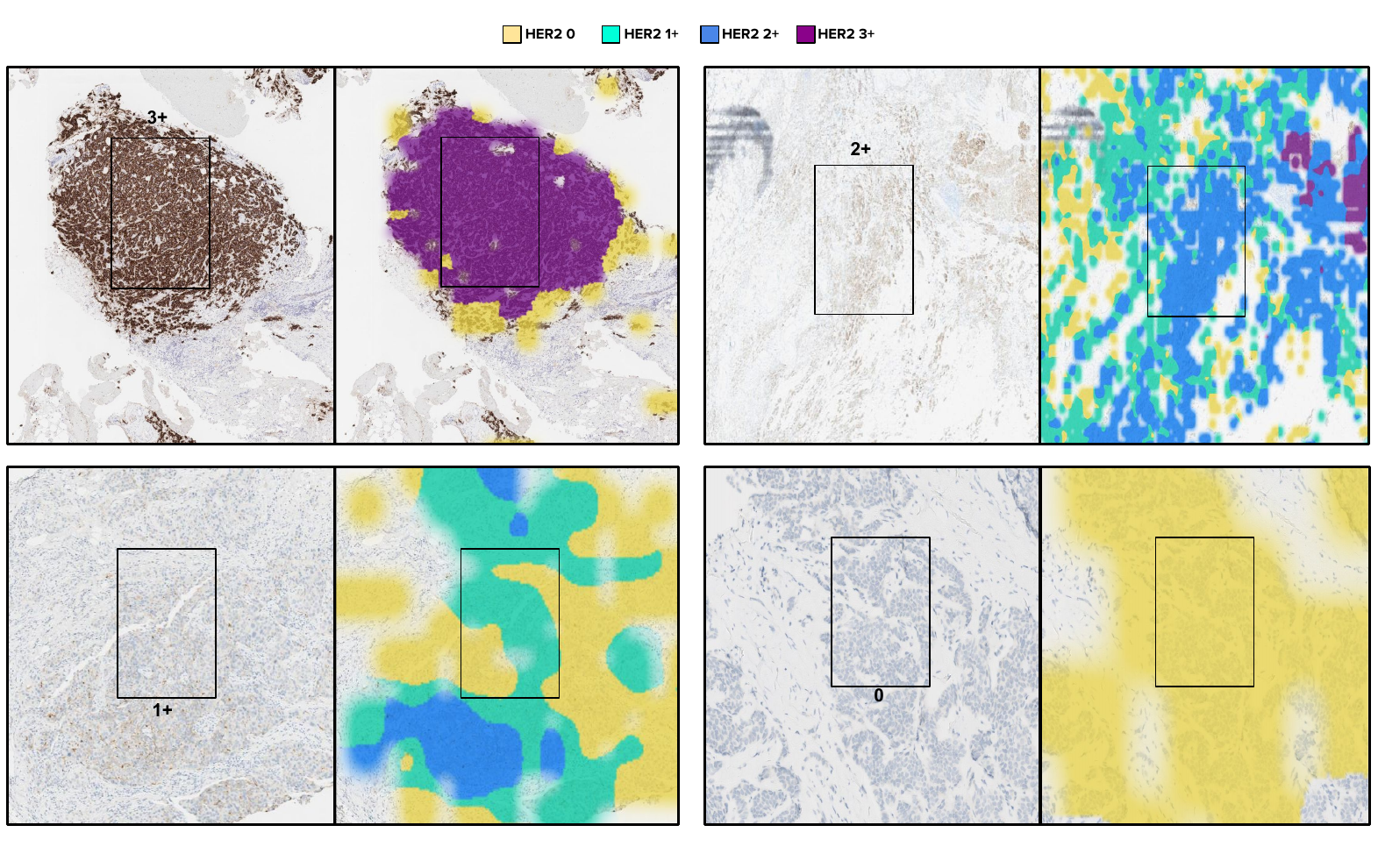}
    \caption{Comparison of AdditiveMIL heatmaps (right) from PLUTO against ground truth ROIs (left) for HER2 scores $3+$, $2+$, $1+$, $0$.  There is considerable alignment between the ground truth ROIs and the PLUTO model's region-level predictions, indicating that the model is learning  biologically-relevant features when making slide-level predictions.}
    \label{fig:additivemil_heatmaps}
\end{figure*}

In the case of NSCLC, the OOD slides exhibit distinct characteristics compared to the ID TCGA slides. As a result, all models experience a decrease in performance. However, PLUTO experiences a comparatively smaller drop and emerges as the top performer among them. On HER2, the performance between ID and OOD datasets is more comparable, and PLUTO performs better compared to all backbones, including the fine-tuned ShuffleNet.
We use AdditiveMIL \cite{javed2022additive} to generate heatmaps to identify regions on the slide corresponding to different HER2 scores. We show examples comparing the heatmaps with ground truth ROIs in Figure \ref{fig:additivemil_heatmaps}.

\begin{table*}
    \centering

\resizebox{\textwidth}{!}{
    \begin{tabular}{cccccccc}
     \thickhline
     
         Model& Dataset& Patch Size&  Tuning&  In-domain F1&  In-domain AUROC&  OOD F1& OOD AUROC\\
         \hline
        \hline

  PLUTO& NSCLC & $16$ & Frozen& $\mathbf{90.2 (1.9)}$& $\mathbf{94.0 (1.6)}$& $\mathbf{86.1 (2.8)}$&$\mathbf{91.2 (2.5)}$\\

Meta-DINOv2 ViT-S& NSCLC & $14$ & Frozen& $88.6 (2.0)$& $92.0 (1.9)$& $72.1 (4.1)$& $81.9 (3.8)$\\

ShuffleNet& NSCLC & - & Frozen& $83.6 (2.4)$& $90.1 (2.0)$& $72.2 (4.2)$&$83.5 (3.5)$\\
ShuffleNet& NSCLC & - & Fine-tuned& $88.1 (2.2)$& $93.9(1.5)$& $42.5 (8.0)$& $90.8 (2.1)$\\
        \hline
         PLUTO& HER2 &  $16$ & Frozen& $ \mathbf{71.5 (2.8)}$&  $\mathbf{89.5 (1.5)}$&  $\mathbf{71.0 (3.0)}$&$ \mathbf{93.7 (1.1)}$ \\
         Meta-DINOv2 ViT-S& HER2&  $14$ & Frozen&  $56.4 (3.2)$&  $83.4 (1.7)$&  $57.2 (3.5)$& $88.7 (1.3)$\\
         ShuffleNet& HER2&  - & Frozen& $51.3 (3.2)$ & $84.4 (1.7)$ &  $46.6 (3.0)$& $86.5 (1.5)$\\
         ShuffleNet& HER2&  - & Fine-tuned& $62.9 (3.1)$ & $87.2 (1.5)$ & $66.3 (3.4)$& $91.6 (1.3)$\\
         \thickhline
    \end{tabular}}
 \caption{Performance of MIL models with different ViT- and CNN-based featurizers on NSCLC subtyping and HER2 scoring tasks. The mean and standard deviation across $1,000$ bootstrapped runs are reported. We note that MIL models that use our frozen PLUTO model as a featurizer tend to outperform models with both frozen and fine-tuned CNN backbones (ShuffleNet), Imagenet-pre-trained ViT backbones, and similar pathology pre-trained ViTs where applicable (NSCLC subtyping). This is especially evident in OOD performance, highlighting the robustness of PLUTO's embeddings. }
\label{table:MIL}
\end{table*}

\subsection{Tissue-level Results}

\subsubsection{Tile Classification: Public Datasets}

We used two publicly available datasets: CRC-100K \cite{NCTCRC} and Camelyon17-WILDS \cite{camelyon17, camelyon17_wilds}. 
The CRC-100K dataset consists of $107,180$ images ($224\times224$ at $0.5$ mpp) of human colorectal cancer (CRC) and normal tissue extracted from $136$ H\&E histopathology WSIs from the NCT Biobank and the UMM pathology archive, classified into one of nine tissue classes. The training set consists of $100,000$ images (referred as NCT-CRC-HE-100K) and the evaluation set consists of $7,180$ images (referred as CRC-VAL-HE-7K). Performance was measured using accuracy (Acc.) and balanced accuracy (Bal. Acc.) and results are shown in Table \ref{table:public_benchmarks}.

\begin{table*}
\centering
\resizebox{0.95\textwidth}{!}{
\begin{tabular}{c c c c c}
 \thickhline
 Model & Adaptation Head & Benchmark Name & \multicolumn{2}{c}{Metrics}\\
  \hline
  \hline
  & & & Acc. & Bal. Acc. \\
 \textbf{PLUTO}  & \textbf{Linear Head} &\textbf{H\&E CRC-100K}  & $\textbf{96.6}$ & $\textbf{95.3}$ \\
   ResNet50* & N/A & H\&E CRC-100K & $94.7$ & N/A \\
   \hline
  & & & Acc. & Bal. Acc. \\
   \textbf{PLUTO}  & \textbf{Linear Head} & \textbf{Camelyon17-WILDS} &  $\textbf{96.2}$ & - \\
    DenseNet-121* & N/A & Camelyon17-WILDS & $70.3$ & - \\
  \hline
  & & & DICE & IoU \\
 \textbf{PLUTO}  & \textbf{Mask2Former} & \textbf{GlaS}  & $\textbf{91.2}$ & $\textbf{84.5}$  \\
 PLUTO  & Mask R-CNN & GlaS  & 88.0 & 79.6  \\
 UNet* & N/A & GlaS  & 85.5 & 74.8 \\
  \hline
  & & & bPQ & mPQ \\
   \textbf{PLUTO}  & \textbf{HoverNet} & \textbf{PanNuke}  & $\textbf{67.1}$ & $\textbf{47.7}$  \\
   PLUTO  & Mask R-CNN & PanNuke  & 58.6 & -  \\
 ResNet50 + Mask R-CNN* \cite{shui2023unleashing}  & N/A & PanNuke  & 55.3 & 36.9  \\
\thickhline
\multicolumn{5}{l}{\small *Fully Supervised Baseline Model} \\
\end{tabular}}
\caption{Summary of PLUTO performance across public datasets. Details of the tasks associated with the datasets are outlined in Table \ref{table:all_benchmarks_overview}. We perform the tile classification task on CRC-100K and Camelyon17-WILDS datasets by linear probing the PLUTO embeddings while keeping the backbone frozen. PLUTO,  while significantly smaller, achieves strong performance and is competitive with the best performing models that have been reported for these two datasets, highlighting the effectiveness of diverse pre-training data for enhancing robustness. We perform the gland segmentation and nuclei segmentation tasks on the GlaS and PanNuke datasets, respectively, by adapting PLUTO through multiple adaptation strategies while keeping the backbone frozen. PLUTO achieves state-of-the-art performance on gland segmentation, outperforming other fully supervised segmentation frameworks. PLUTO beats fine-tuned backbone baselines of comparable size on the PanNuke dataset, and is competitive with significantly larger fine-tuned backbones that have been reported for PanNuke.}
\label{table:public_benchmarks}
\end{table*}

The Camelyon17-WILDS dataset contains $455,954$ images ($96\times96$ pixels at $1$ mpp, downsampled from $0.25$ mpp slides) from $50$ WSIs of breast cancer metastases in lymph node sections from five different hospitals. The task is a binary classification of whether the central $32\times32$ region contains tumor tissue. The training set consists of $302,436$ tiles from $30$ WSIs from three hospitals, the ID validation set of $33,560$ from the same $30$ WSIs, the OOD validation set of $34,904$ from $10$ WSIs from the fourth hospital, and the OOD test set of $85,054$ from $10$ WSIs from the fifth hospital. Each split has a $50/50$ class balance. Performance was evaluated using accuracy in the OOD test set, measuring robustness to shifts across hospitals (Table \ref{table:public_benchmarks}).

The results of our embedding probing on these external tissue classification datasets are summarized in Table \ref{table:public_benchmarks}. PLUTO obtained competitive performance in both the CRC-100K dataset and Camelyon17-WILDS dataset with the best performing models that have been reported on these datasets, despite having significantly fewer parameters and a smaller---but highly diverse---pre-training dataset.

\subsubsection{Tile Classification: Proprietary Datasets}
\label{sec:tissue_tile_class}

We probe the CLS-token and patch-token embeddings on tile-level tissue classification in a broad set of indications and stains. All labels are derived from board-certified pathologists, and substances are chosen to capture the most relevant biology in those indications. In particular, we consider the following to test the adaptability of PLUTO across diverse diseases, stains, and organs:
\begin{itemize}
\item Four-class H\&E Pan-Oncology Tissue: cancer, necrosis, cancer-associated stroma, normal tissue
\item Four-class IHC Pan-Oncology Tissue: cancer, necrosis, cancer-associated stroma, normal tissue
\item $10$-class H\&E Inflammatory Bowel Disease (IBD) Tissue: crypt abscess, inter-gland lumen, infiltrated epithelium, normal tissue, other tissue, blood vessel, granulation tissue, erosion or ulceration, lamina propria, muscularis mucosa.
\end{itemize}

\begin{figure*}
    \centering
    \includegraphics[width=\linewidth]{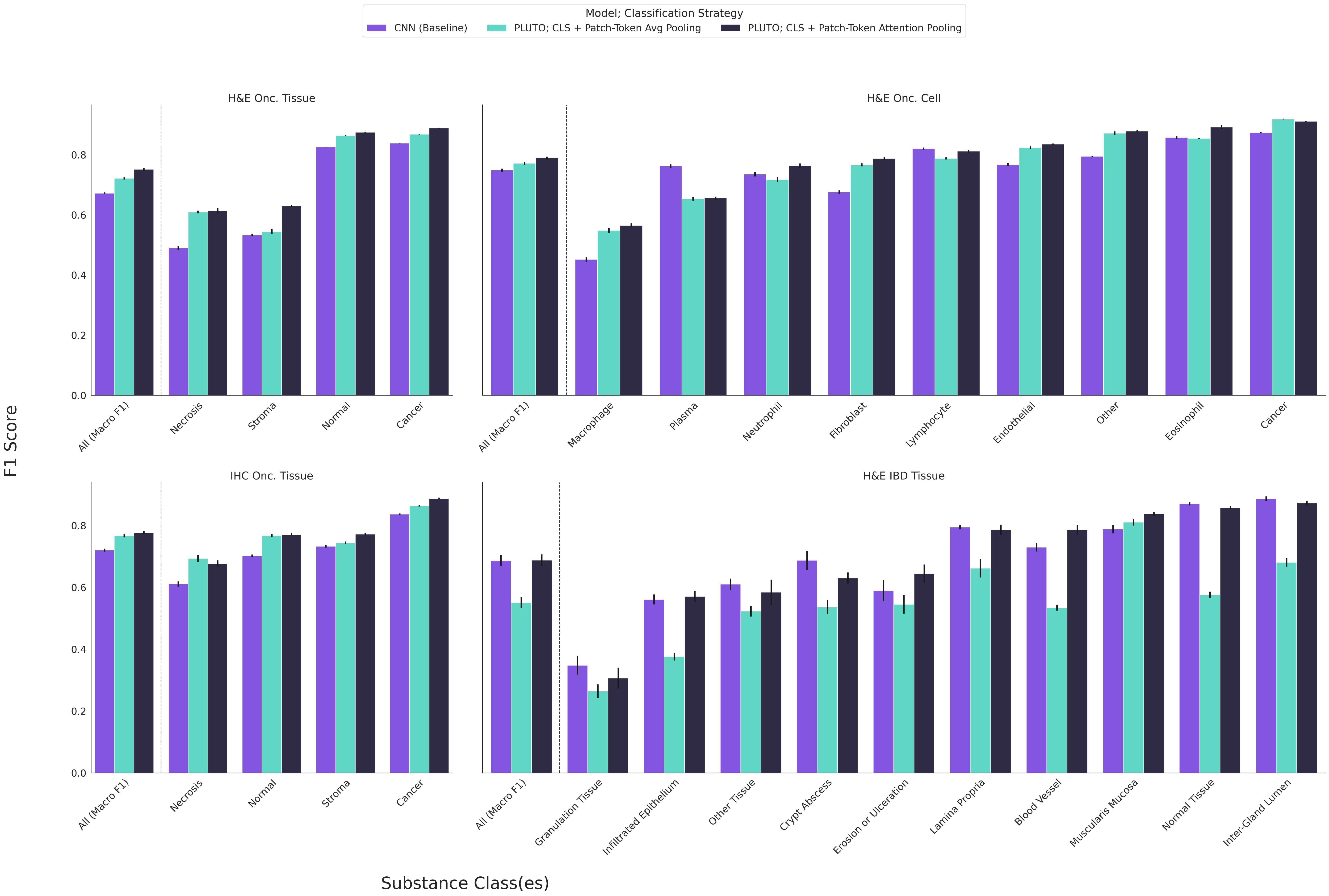}
    \caption{Linear and Attention Probing performance of PLUTO on proprietary tile classification benchmarks at the substance level. For each model, the left-most bar highlights the macro-F1 score on the dataset.  PLUTO outperforms a fully supervised CNN baseline.  Attentive pooling of patch-tokens provides more flexible adaptation and has the best performance across datasets. For the Oncology cell classification task, patch-token information from the central window is needed to capture context to classify the cell at the center pixel, so average pooling and attention pooling perform comparably. While patch embeddings do contain relevant information for the more complicated task of IBD tissue classification, performance significantly improves on applying attention pooling on top of them.}
    \label{fig:patch_class_internal_bar}
\end{figure*}

 A FlexiViT patch size of $16\times16$ was used for tissue tasks, while $8\times8$ is used for cell classification. We begin by treating the tissue task as a simple tile classification problem and use an MLP on top of the CLS tokens for probing. Along with simple probing, we use a variant of attentive probing \cite{chen2023context}, where the CLS embedding is augmented by concatenating an attention-pooled summary of the patch-tokens. The attention pooling allows the classifier to learn relevant context from across the entire image, and results in a performance improvement over only using the CLS embedding. Supplementary Tables \ref{table:center_pixel_tissue}, \ref{table:center_pixel_cell} show macro-F1 scores for each model and embedding pooling set-up.

Comparing Oncology and IBD Tissue results demonstrates that the best adaptation approach is dependent on the target task. On a four-class tissue problem, the CLS token holds the vast majority of the necessary signal, and adding the learned attention pooling results in less significant benefits. However, in $10$-class IBD Tissue, the classifier struggles when relying only on the CLS token and must instead learn to attend to selective patches.

To contextualize the results, we train a ResNet-style fully supervised CNN model from scratch on the same dataset. We demonstrate that by using PLUTO in combination with a small amount of labeled data and an optimal adaptation approach, we out-perform the baseline fully-supervised model on Oncology indications and achieve statistically equivalent performance on IBD Tissue Classification. We reason that the level of detail necessary to distinguish morphology among IBD classes warrants further exploration into optimal light-weight methods for extracting information from frozen embeddings. Figure \ref{fig:patch_class_internal_bar} captures the class-level and aggregated performance for all datasets.

\subsubsection{Gland Segmentation: Public Datasets}
Gland morphology is used in gastrointestinal (GI) tract pathology. 
The architectural appearance of glands is vital for cancer grading in colorectal carcinoma \cite{compton2000updated}. 
In ulcerative colitis or IBD, changes in crypt architecture (density, morphology, etc.) are a core component of the Geboes scoring system \cite{graham2022screening}. 
Outside of the GI tract, gland differentiation is important for diagnosis and grading in breast and other cancers \cite{cserni2020grading}.
We evaluated the performance on the GlaS \cite{glas} dataset, which consists of $85$ images for training and $80$ images for testing for a total of $165$ images derived from $16$ H\&E-stained sections of stage T3 or T42 colorectal adenocarcinoma. These slides were scanned using a Zeiss MIRAX MIDI Slide Scanner with a resolution of $0.465$ mpp and varying image sizes (most commonly $775\times522$). Results are shown in Table \ref{table:public_benchmarks}. Performance was measured using dice coefficient (Dice) and Intersection over Union (IoU) in the test set. 
To the best of our knowledge, this is the first pathology FM used in a gland segmentation task. Results show state-of-the-art performance, with PLUTO outperforming other fully-supervised segmentation frameworks. 
This underscores the ability of our proposed approach to successfully adapt to a new tissue-level segmentation task with limited labeled data. 

\subsubsection{Gland Segmentation: Proprietary Datasets}
\label{sec:gland_int}

\begin{figure*}
    \includegraphics[width=\linewidth]{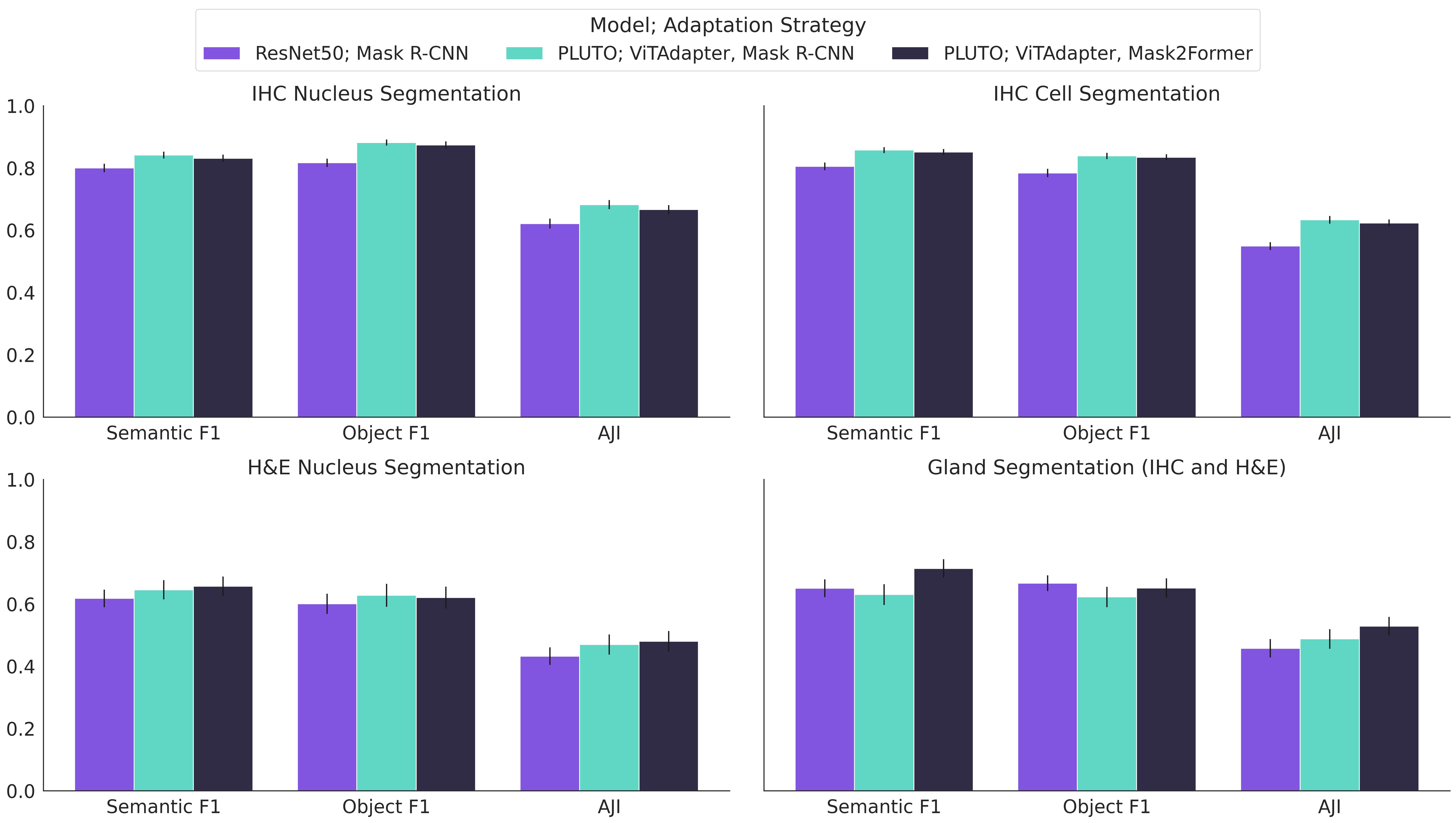}
    \caption{Instance segmentation performance on our proprietary datasets across IHC nuclei segmentation, IHC cell segmentation, H\&E nuclei segmentation, and gland segmentation (H\&E and IHC). Point estimates show the average metric across the set of tiles, and error bars represent a 95\% confidence interval. The baseline performance is established with a fully supervised Mask R-CNN model with a ResNet50 backbone. The PLUTO ViTAdapter + Mask R-CNN and ViTAdpter + Mask2Former adaptation heads are fine-tuned while the backbone is kept frozen.   We note that our flexible patch size approach allowed us to change the patch size across tasks while leaving the PLUTO weights frozen.}
    \label{fig:internal_instance_seg}
\end{figure*}

We also curate an internal instance segmentation dataset to segment glands across stains, organs, and disease areas. We discuss the characteristics of the dataset in Table \ref{tab:gland_seg_data} and results in Figure \ref{fig:internal_instance_seg} with qualitative comparisons in Figure \ref{fig:segmentation_heatmaps}. We train on samples of size $768\times768$ at a resolution of $1$ mpp using our proposed adaptation approaches using a patch size of $16$. Both of our adaptation approaches beat the Mask R-CNN baseline using a frozen ResNet50 backbone, with the Mask2Former adaptation head providing the highest performance across all metrics. We further experiment with a patch size of $32$ and notice a very minor performance drop (1\%) but significantly higher inference throughput.  We notice that Mask2Former works better on glands where the objects are not close to convex-like shapes; we hypothesize that the query based segmentation mechanism is helping the network, and we aim to explore this further in future work.

\subsection{Cellular- and Subcellular-level Results}
\subsubsection{Tile Classification: Proprietary Datasets}
Tile classification is extended to a nine-class H\&E Oncology cell benchmark: Macrophages, Plasma Cells, Neutrophils, Fibroblasts, Lymphocytes, Endothelial Cells, Eosinophils, Cancer Cells, and Other Cells. The label for each tile is derived from the cell class present at  the center pixel of the tile. Because the CLS token does not capture the granular information needed for cell classification, we concatenate it with an average pooling of the four central patch token embeddings. We then extend the approach to Attention Pooling in the same fashion as Section \ref{sec:tissue_tile_class} by allowing the learned attention layer to see all patch tokens in the tile. The result is shown alongside the tissue classification tasks in Figure \ref{fig:patch_class_internal_bar}, and full ablations are in Supplementary Tables \ref{table:center_pixel_tissue},  \ref{table:center_pixel_cell}. The context in the center four patch tokens provides sufficient signal for cell classification, but learned attention pooling on the entire tile achieves slightly better results on almost all substances.

\subsubsection{Nuclei and Cell Segmentation: Public Datasets}

Instance segmentation is a popular approach for nucleus, cytoplasm, and cell quantification on H\&E and IHC-stained WSIs \cite{pathainuclei}.

We evaluated the performance of our adaptation strategies on the PanNuke \cite{pannuke} dataset. It consists of $481$ visual fields across $19$ different tissue types from WSIs from TCGA and a local hospital, with a total of $189,744$ exhaustive nuclei labels categorized into five classes. The visual fields were randomly sampled from more than $20,000$ WSIs that were scanned at either $40\times$ or $20\times$ and re-sized to $40\times$.  Following the original publishers of this dataset, we report binary panoptic quality (bPQ) and multi-class panoptic quality (mPQ) in Table \ref{table:public_benchmarks}. For the ablation study comparing different adaptation heads, experiments were conducted in the binary configuration where nuclei were not classified, and therefore only bPQ is reported. The experimental setup used an inference patch size of $16$ for the HoverNet architecture due to its design. All conducted experiments were thus performed using this specified patch size.

Results demonstrated that despite its significantly smaller architecture, PLUTO (without fine-tuning) achieved nuclei segmentation performance comparable to state-of-the-art with much larger models that have been reported on this dataset. In the binary setup, the comparison of the HoverNet adaptation head to the Mask R-CNN variation highlighted the advantage of using specialized adaptation architectures designed in favor of nuclei segmentation. 

\subsubsection{Nuclei and Cell Segmentation: Proprietary Datasets} 
\begin{figure*}
    \includegraphics[width=\linewidth]{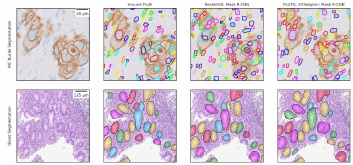}
    \caption{Comparison of instance segmentation masks across models at the subcellular level on an IHC slide (top row, Nuclei Segmentation) and at the tissue level on an H\&E slide (bottom row, Gland Segmentation) on proprietary datasets. These tiles visualize a representative, smaller field of view within larger labeled tiles where the quantitative performance, characterized here by AJI \cite{7872382}, aligns with dataset-level metrics. IHC Nuclei Segmentation in the whole tile using \textbf{ResNet50; Mask R-CNN} reaches $0.448$ AJI and using \textbf{PLUTO, ViTAdapter; Mask R-CNN} reaches $0.527$ AJI. Gland Segmentation on the whole tile using \textbf{ResNet50; Mask R-CNN} reaches $0.445$ AJI and using \textbf{PLUTO, ViTAdapter; Mask R-CNN} reaches $0.450$ AJI. In both cases, PLUTO outperforms the \textbf{ResNet50; Mask R-CNN} baseline.}
    \label{fig:segmentation_heatmaps}
\end{figure*}

IHC \cite{ihc} is a widely used tool in pathology for disease diagnosis and subtyping, cell classification, and quantification of protein abundance. 
IHC staining utilizes antibodies targeted against certain antigens in specific tissues and cells to directly quantify protein abundance, and often its localization within the cell (e.g., membranous, cytoplasmic, or nuclear staining). Thus, especially in IHC, it is important not only to identify individual cells and nuclei but also to precisely delineate their boundaries. We curate internal datasets for IHC Cell Segmentation, IHC and H\&E Nuclei Segmentation. We describe the characteristics of our proprietary nuclei and cell segmentation datasets in Tables \ref{tab:nucseg} and \ref{tab:cellseg} in the Supplementary Section. We compare Mask R-CNN with ResNet50 and ViT backbones using patch size $8$, and compare Mask R-CNN with a Mask2Former adaptation head. The results from these experiments on IHC Nuclei and Cell Segmentation and H\&E Nuclei Segmentation are summarized in Figure \ref{fig:internal_instance_seg} with qualitative comparisons in Figure \ref{fig:segmentation_heatmaps}. We train on samples of size $384\times384$ at a resolution of $0.25$ mpp. We find that our adaptation approaches beat the Mask R-CNN baseline using a frozen ResNet50 backbone.

\subsection{Deployability} 
\label{sec:deploy}

Given the increasing adoption of ML in digital pathology for clinical and diagnostic use-cases, there is a need for ML algorithms which are robust and can be deployed at scale to address patient and clinician needs. Real-world deployment of ML algorithms needs careful consideration of factors like deployment throughput, algorithm latency for end-users, and cost of deployment. While FMs can enable new capabilities and improve generalization, they are often orders of magnitude larger than task-specific models, and deploying them on WSIs can be significantly more expensive and time-consuming. 

\begin{figure}

    \includegraphics[width=\linewidth]{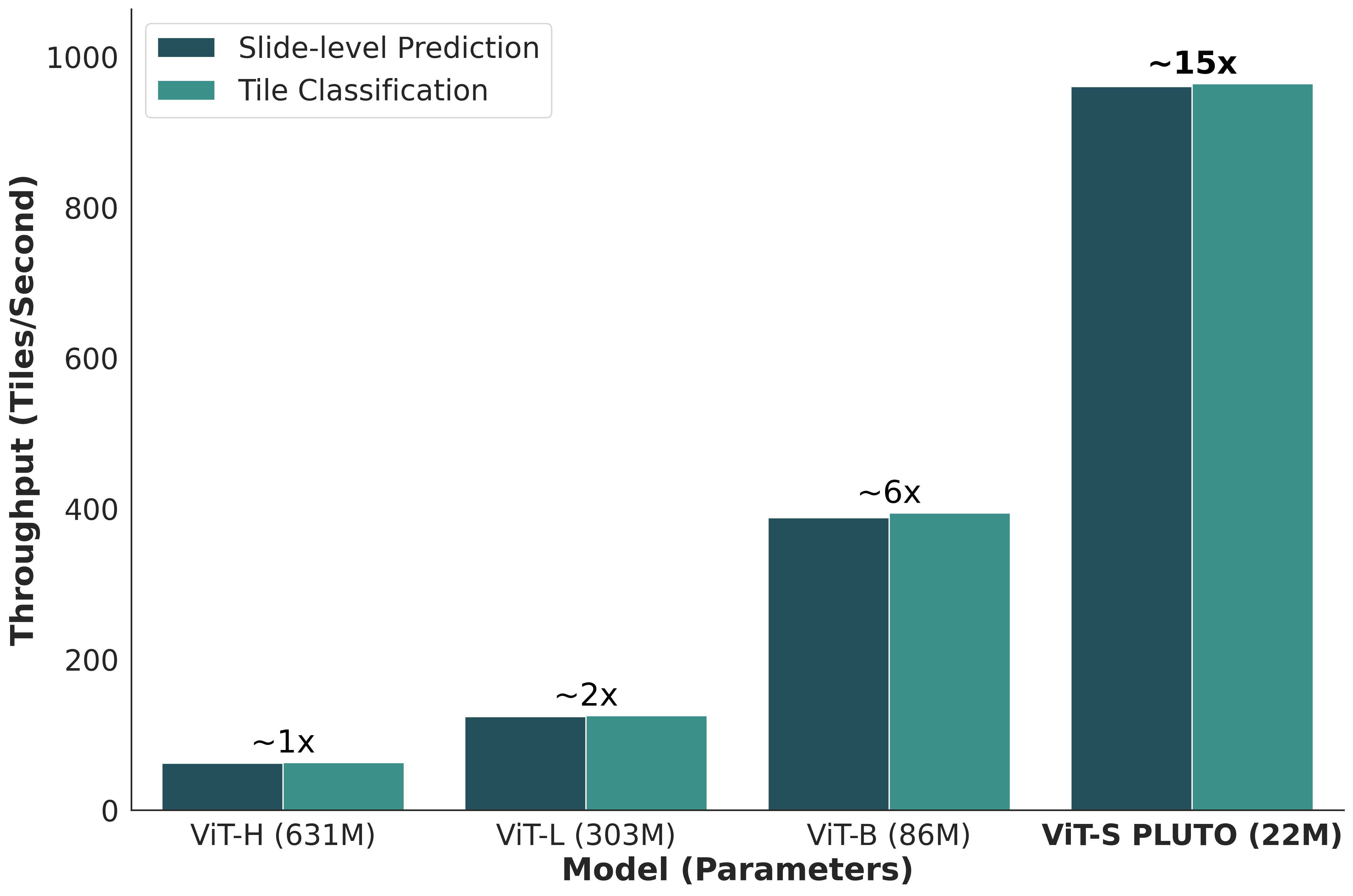}
    \caption{Throughput (tiles/sec) of models for tile-level and slide-level classification tasks with various backbones using patch size $16$ with a tile size $224\times224$. We use linear probes and AdditiveMIL classifiers as adaptation heads respectively for the tile and slide-level classification tasks. Notable pathology FMs use ViT-H \cite{virchow}, ViT-L \cite{uni} \cite{rudolfv} and ViT-B \cite{phikon}.\\}
    \label{fig:throughput}
\end{figure}

To develop a backbone which is performant and robust while being suitable for deployment at scale, we make two key design choices for PLUTO. We choose the ViT architecture ViT-S as our backbone, as it is light-weight while still having enough model capacity to be performant at different tasks. Additionally, the FlexiViT architecture enables customization of the patch size, which not only enables selecting the optimal patch size for a downstream task, but can also be used to improve model throughput as the patch size controls the sequence length. To illustrate the efficiency of PLUTO, we compare the throughput efficiency of various ViT backbones (ViT-S, ViTB, ViT-L, ViT-H) for two common pathology tasks: tile classification and slide-level prediction. We fix the task-specific adaptation heads---linear layer for patch-classification, and AdditiveMIL for slide-level prediction---while varying the backbone, and measure the throughput in tiles/second with a tile size of $224\times224$ and a patch size of $16$.  We note that we have not applied any inference-specific optimizations in this setup. We use the same data-loading pipeline and hardware (A40 GPU) for all the backbones. As seen in Figure \ref{fig:throughput}, for both the tasks, ViT-S is around $2.5\times$ faster than ViT-B, $7.5\times$ faster than ViT-L, and $15\times$ faster than ViT-H.

\section{Summary and Future Work}
We present in this paper PLUTO: a competitive state-of-the-art pathology Foundation Model based on a light-weight ViT that is pre-trained on a diverse dataset from over $50$ distinct sites consisting of over $195$ million unique image tiles across four resolutions ($0.25$, $0.5$, $1$, and $2$ mpp) sampled from $158,852$ WSIs, over $16$ tissue groups, and $28$ disease areas.  We implement the PLUTO pre-training scheme by modifying the DINOv2 \cite{dinov2} self-supervised training strategy to incorporate different patch sizes and WSI tile resolutions through extending the FlexiViT \cite{flexivit} framework. We further modify the DINOv2 loss function by adding a MAE \cite{mae} objective term and introducing a Fourier loss term that encourages capturing low-frequency and high-frequency features. 

PLUTO is designed to take advantage of the multi-scale nature of WSIs and provide informative representations across biological scales. We have quantified the performance of our self-supervised backbone on a variety of adaptation tasks across biological scales, namely:

\begin{itemize}
    \item \textbf{Slide-level classification} We adapt PLUTO to slide-level scoring tasks with multiple instance learning (MIL) and achieve superior OOD performance on NSCLC subtyping and HER2 scoring.
    \item \textbf{Tile Classification: Tissue} We adapt PLUTO to a nine-class tissue classification task on the NCT-CRC dataset \cite{NCTCRC} and to a binary breast cancer classification task on the Camelyon17-WILDS \cite{camelyon17_wilds} dataset. We demonstrate results competitive with state-of-the art methods that have been reported on these two datasets. We additionally quantify the quality of these representations (Section \ref{sec:tissue_tile_class}) on a variety of tasks on multiple proprietary datasets, namely on a four-class tissue classification task across oncology indications on H\&E and IHC-stained WSIs, and on a $10$-class tissue classification task for IBD. PLUTO matches or outperforms a supervised CNN baseline in these datasets.
    \item \textbf{Instance Segmentation: Gland Segmentation} We adapt PLUTO  with Mask R-CNN and Mask2Former adaptation heads respectively to identify glands through instance segmentation. We quantify the performance of this approach with two datasets: the public GlaS \cite{glas} dataset and our proprietary gland annotations. On the GlaS dataset, we achieve state-of-the-art performance compared to public strongly-supervised benchmarks. On our internal dataset, PLUTO performs superior to ResNet50 backbone using our Mask2Former approach. 
    \item \textbf{Tile Classification: Cells and Nuclei}  We quantify the quality of PLUTO embeddings on a tile classification task of nine-class cell classification across H\&E stained slides, where we also outperform a supervised CNN baseline.
    \item \textbf{Instance Segmentation: Cells and Nuclei} We adapt PLUTO with Mask R-CNN and Mask2Former adaptation heads respectively to identify cells and nuclei through instance segmentation. We have measured the performance of this approach with four segmentation datasets -- one public (PanNuke \cite{pannuke}) and three proprietary. We achieve a higher performance with our PLUTO Mask R-CNN and Mask2Former adaptation heads on our proprietary datasets than baselines we have established with a ResNet50 backbone. We additionally adapt PLUTO with the HoverNet architecture \cite{hovernet} to identify nuclei on the PanNuke dataset, where we achieve performance competitive to state-of-the-art methods that have been reported on PanNuke.
\end{itemize}

Our work also demonstrates the importance of data diversity, along with the incorporation of biological priors in the construction of pre-training datasets and the design of the model architecture for large-scale self-supervised models. PLUTO was trained with the ViT-S architecture consisting of $22$ million parameters, which is significantly fewer than the number of parameters of the other pathology FMs discussed in Section \ref{sec:intro}. Despite its smaller size, we demonstrate competitive OOD performance on public benchmarks spanning tile classification, WSI-level prediction, and instance segmentation.  Additionally, the smaller model size of PLUTO and the flexibility to customize the patch size enables a significantly faster inference time in comparison to other pathology FMs, which makes PLUTO well-suited for deployments at scale. Our results hint that dataset diversity could be an additional empirical scaling law that complements the well-established scaling laws of dataset size, model size and amount of compute \cite{openaiscale}. Further research is needed, however, to establish empirical scaling laws for digital pathology in order to understand the impact of dataset diversity, dataset size, model size, and amount of compute on model performance, robustness, and deployability. 

A multi-scale pathology FM that learns strong representations across biological scales and  demonstrates robust performance on tasks across these scales will enable computational pathology to shift from bespoke single-task models towards more general AI models that can address long-standing problems with diverse applications in pathology. We hope that our efforts with PLUTO further motivate building high-performing, deployable FMs; drive FM adoption in pathology; and serve real-world translational research and clinical applications.

{
    \small
    \bibliographystyle{ieeenat_fullname}
    \bibliography{main}
}
\setcounter{section}{0}

\def\thickhline{\noalign{\hrule height0.8pt}}
\maketitlesupplementary

\begin{table}
\centering
\begin{tabular}{ c c }
 \thickhline
 Organ& Number of Slides \\
  \hline
Breast& $19,936$ \\
Lower GI& $19,185$ \\
Lung& $15,418$ \\
Urinary System& $14,167$ \\
Skin& $13,463$ \\
Prostate& $12,955$ \\
Liver and Biliary Tract& $12,840$ \\
Lymphatic System& $9,586$ \\
Female Genital Tract& $9,297$ \\
Upper GI& $7,643$ \\
Unknown / Other& $6,140$ \\
Pancreas& $5,097$ \\
Head and Neck& $5,030$ \\
Brain& $3,884$ \\
Bone and Connective Tissue& $2,297$ \\
Endocrine& $1,547$ \\
Male Genital Tract& $367$ \\
 \thickhline
\end{tabular}
\caption{Organ distribution of pre-training dataset.}
\label{table:organ}
\end{table}

\begin{table}
\centering
\begin{tabular}{ c c }
 \thickhline
 Scanner& Number of Slides \\
 \hline
Aperio GT 450& $45,476$ \\
Aperio AT2& $43,905$ \\
Aperio NOS& $30,107$ \\
Philips UFS& $12,317$ \\
Ventana DP 200& $11,686$ \\
Unknown / Other& $11,272$ \\
Hamamatsu S360 C13220& $2,318$ \\
Hamamatsu XR C12000& $1,144$ \\
3D Histech Pannoramic Scan II& $243$ \\
Hamamatsu S210 C13239& $143$ \\
Aperio AT Turbo& $138$ \\
Hamamatsu 2.0-HT& $103$ \\
 \thickhline
\end{tabular}
\caption{Scanner distribution of pre-training dataset.}
\label{table:scanner}
\end{table}

\begin{table}
\centering
\begin{tabular}{ c c }
 \thickhline
 Objective Magnification& Number of Slides \\
 \hline
 $20\times$& $35,563$ \\
 $40\times$& $123,289$ \\
 \thickhline
\end{tabular}
\caption{Objective magnification distribution of pre-training dataset.}
\label{table:objective}
\end{table}

\begin{table}
\centering
\begin{tabular}{ c c }
 \thickhline
 Disease& Number of Slides \\
 \hline
Breast Cancer& $20,062$ \\
Lung Cancer& $15,905$ \\
Prostate Cancer& $13,122$ \\
Non-neoplastic Lower GI& $12,699$ \\
Lymphoma& $8,482$ \\
Non-neoplastic Liver& $8,317$ \\
Other Skin Lesion& $7,149$ \\
Malignant NOS / Other& $6,539$ \\
Urothelial Cancer& $6,477$ \\
Unknown / Other& $5,753$ \\
Gastric Cancer& $5,580$ \\
Pancreatic Cancer& $5,144$ \\
Melanoma& $4,959$ \\
Head and Neck Cancer& $4,746$ \\
Renal Cell Cancer& $4,658$ \\
Endometrial Cancer& $4,621$ \\
Colorectal Cancer& $4,134$ \\
Glioma& $3,562$ \\
Ovarian Cancer& $3,159$ \\
Hepatobiliary Cancer& $2,562$ \\
Non-neoplastic Other& $2,514$ \\
Non-neoplastic Upper GI& $1,973$ \\
Non-neoplastic Kidney& $1,587$ \\
Cervical Cancer& $1,265$ \\
Thyroid Cancer& $1,023$ \\
Sarcoma& $979$ \\
Basal Cell Cancer& $755$ \\
Bladder Cancer& $684$ \\
Multiple Myeloma& $442$ \\
 \thickhline
\end{tabular}
\caption{Disease distribution of pre-training dataset.}
\label{table:disease}
\end{table}

\begin{table}
\centering
\begin{tabular}{ c c }
 \thickhline
 Stain& Number of Slides \\
 \hline
H\&E (FFPE)& $89,829$ \\
IHC& $53,498$ \\
H\&E (Frozen)& $7,091$ \\
Special Stain& $4,856$ \\
Unknown / Other& $3,578$ \\
 \thickhline
\end{tabular}
\caption{Stain distribution of pre-training dataset. Special stains include trichrome, iron, periodic acid-Schiff, sirius red, Alcian blue, and Grocott-Gomori methenamine silver.}
\label{table:stain}
\end{table}

\begin{table*}
\centering
\resizebox{\textwidth}{!}{
\begin{tabular}{ c c c c c c } 
 \thickhline
 Model & Data Source & Dataset Size & Backbone & Model Size & Method \\
 \hline
 Virchow \cite{virchow} & Proprietary (MSKCC) & $1.5$M WSIs / $2$B tiles & ViT-H & $632$M & DINOv2 \\
 UNI \cite{uni} & Proprietary (Mass-100K) & $100$K WSIs / $100$M tiles & ViT-L & $307$M & DINOv2 \\
 RudolfV \cite{rudolfv} & TCGA + Proprietary & $103$K WSIs / $750$M tiles & ViT-L & $307$M & DINOv2 \\ 
 Phikon \cite{phikon}& TCGA & $6$K WSIs / $43$M tiles & ViT-B & $86$M & iBOT \\ 
 Kang \textit{et al.} \cite{lunit} & TCGA + Proprietary & $37$K WSIs / $33$M tiles & ViT-S & $22$M & DINOv1 \\
 PLUTO (this work) & TCGA + Proprietary (PathAI) & $158$K WSIs / $195$M tiles & FlexiViT-S & $22$M & Modified DINOv2 \\
 \thickhline
\end{tabular}
}
\caption{Survey of recent pathology foundation models.  Our proprietary dataset consists of a diverse spectrum of histology stains, scanners, biological objects, and tissue regions across resolution scales sourced from over $50$ sites. We customized the DINOv2 training procedure by including a Masked Autoencoder (MAE) \cite{mae} loss term, as well a Fourier loss term that decomposes the WSI tile to low- and high-frequency components.}
\label{table:path_fms}
\end{table*}

\begin{figure*}[h!]
    \includegraphics[width=\linewidth]{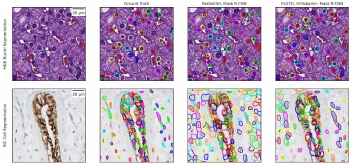}
    \caption{Comparison of segmentation masks across models at the subcellular level on an H\&E slide (top row, Nuclei Segmentation) and at the cellular level on an IHC slide (bottom row, Cell Segmentation) on proprietary datasets. These tiles visualize a representative, smaller field of view within larger labeled tiles where the quantitative performance, characterized here by AJI \cite{7872382}, aligns with dataset-level metrics. H\&E Nuclei Segmentation in the whole tile using \textbf{ResNet50; Mask R-CNN} reaches $0.474$ AJI and using \textbf{PLUTO, ViTAdapter; Mask R-CNN} reaches $0.585$ AJI. IHC Cell Segmentation on the whole tile using \textbf{ResNet50; Mask R-CNN} reaches $0.203$ AJI and using \textbf{PLUTO, ViTAdapter; Mask R-CNN} reaches $0.347$ AJI. In both cases, PLUTO outperforms the \textbf{ResNet50; Mask R-CNN} baseline.}
    \label{fig:sup_segmentation_heatmaps}
\end{figure*}

\begin{table}
    \centering
    \resizebox{\columnwidth}{!}{
    \begin{tabular}{cccc}
    \thickhline
        Disease & Split & Number of Patches & Number of Instances \\ \hline
        Breast & Train & $82$ & $666$\\
        Breast  & Heldout & $26$ & $281$ \\
        CRC & Train & $40$ & $725$ \\
        CRC & Heldout & $16$ & $474$ \\
        GI Tract & Train & $295$ & $9,083$\\
        GI Tract & Heldout & $59$ & $2,814$\\
    \thickhline
    \end{tabular}}
    \caption{Our proprietary Gland segmentation data characterization (includes H\&E and IHC).}
    \label{tab:gland_seg_data}
\end{table}

\begin{table}
    \centering
    \resizebox{\columnwidth}{!}{
    \begin{tabular}{cccc}
    \thickhline
        Disease & Split & Number of Frames & Number of Instances \\ \hline
        Breast & Train & $26$ & $6,362$\\
        
        Breast & Heldout & $17$ & $4,528$ \\
        
        CRC & Train & $31$ & $6,751$ \\
        
        CRC & Heldout & $13$ & $2,630$ \\
        
        Gastric & Train & $29$ & $5,290$ \\
        
        Gastric & Heldout & $6$ & $1,147$ \\
        
        NSCLC & Train & $22$ & $7,129$ \\
        
        NSCLC & Heldout & $19$ & $5,243$ \\
        
        Ovarian & Train & $33$ & $6,249$\\
        
        Ovarian & Heldout & $10$ & $1,933$\\
        
        Esophageal & Train & $17$ & $4,213$\\
        
        Esophageal & Heldout & $2$ & $291$\\
    \thickhline
    \end{tabular}}
    \caption{Our proprietary Cell segmentation data characterization (IHC only).}
    \label{tab:cellseg}
\end{table}

\begin{table*}[t]
\centering
\begin{tabular}{c c c c c} 
 \thickhline
 Task & Model Arch. & Inference Patch Size & Embedding (Pooling Method) & Macro F1 \\ [0.5ex] 
 \hline
 \hline
 
 H\&E Onc. Tissue & PLUTO + MLP & $16$ & CLS (N/A) & $0.720$ \\
  
 H\&E Onc. Tissue & PLUTO + MLP & $16$ & All Patch-Token (Mean) + CLS & $0.722$ \\
 
 H\&E Onc. Tissue & PLUTO + MLP & $16$ & All Patch-Token (Attention) & $0.743$\\
 
 H\&E Onc. Tissue & PLUTO + MLP & $16$ & All Patch-Token (Attention) + CLS & $\mathbf{0.751}$ \\
 
 H\&E Onc. Tissue & CNN (Baseline) & - & - & $0.672$ \\
 \hline
 
 IHC Onc. Tissue & PLUTO + MLP & $16$ & CLS (N/A) & $0.764$ \\
  
 IHC Onc. Tissue & PLUTO + MLP & $16$ & All Patch-Token (Mean) + CLS  & $0.767$ \\
 
 IHC Onc. Tissue & PLUTO + MLP & $16$ & All Patch-Token (Attention)  & $0.768$\\
 
 IHC Onc. Tissue & PLUTO + MLP & $16$ & All Patch-Token (Attention) + CLS & $\mathbf{0.777}$ \\
 
 IHC Onc. Tissue & CNN (Baseline) & - & - & $0.721$ \\
  
 \hline

 H\&E IBD Tissue & PLUTO + MLP & $16$ & CLS (N/A) & $0.544$ \\
  
 H\&E IBD Tissue & PLUTO + MLP & $16$ & All Patch-Token (Mean) + CLS & $0.551$ \\
 
 H\&E IBD Tissue & PLUTO + MLP & $16$ & All Patch-Token (Attention) & $0.658$\\
 
 H\&E IBD Tissue & PLUTO + MLP & $16$ & All Patch-Token (Attention) + CLS & $\mathbf{0.688}$ \\
 
 H\&E IBD Tissue & CNN (Baseline) & - & - & $0.687$\\ 

\thickhline
\end{tabular}
\caption{Performance on tissue tile classification tasks on our proprietary datasets. On oncology tasks, the CLS token contains enough information to perform well. In all cases, adding Attention improves the macro-F1, with the largest gains in IBD Tissue. There, possibly due to the complexity of the task, attention is necessary for the classifier to learn the relevant context for classification. IBD Tissue is also where the CNN Baseline and the PLUTO models perform most similarly.}
\label{table:center_pixel_tissue}
\end{table*}

\begin{table*}[t]
\centering
\begin{tabular}{c c c c c} 
 \thickhline
 Task & Model Arch. & Inference Patch Size & Embedding (Pooling Method) & Macro F1 \\ [0.5ex] 
 \hline
 H\&E Onc. Cell & PLUTO + MLP &  $8$ & Sub Patch-Token (Mean) + CLS & $0.771$ \\
 H\&E Onc. Cell & PLUTO + MLP & $8$ & All Patch-Token (Attention) & $0.760$ \\
 H\&E Onc. Cell & PLUTO + MLP & $8$ & All Patch-Token (Attention) + CLS & $\mathbf{0.789}$ \\
 H\&E Onc. Cell & CNN (Baseline) & - & - & $0.749$ \\
\thickhline
\end{tabular}
\caption{Performance on cell tile classification tasks on our proprietary dataset. We use the mean of the center four patch-token embeddings and concatenate it with the CLS token as the baseline PLUTO embeddings. Adding Attention does not alter performance substantially in this scenario, and all PLUTO models outperform the CNN Baseline.}
\label{table:center_pixel_cell}
\end{table*}

\begin{table}
    \centering
    \resizebox{\columnwidth}{!}{
    \begin{tabular}{cccc}
    \thickhline
        Disease & Split & Number of Frames & Number of Instances \\ \hline
         Breast& Train & $41$ & $4,019$\\
         Breast & Heldout & $23$ & $2,279$ \\ 
         CRC & Train & $23$ & $2,865$\\ 
         CRC & Heldout & $10$ & $1,410$\\ 
         DLBCL & Heldout & $2$ & $912$\\ 
         Gastric & Train & $9$ & $1,098$\\ 
         Gastric & Heldout & $3$ & $394$\\ 
         HCC & Train & $6$ & $2,065$\\ 
         HNSCC & Train & $1$ & $526$\\ 
         Hepatitis & Train &  $3$ & $632$\\ 
         IBD & Train & $1$ & $778$\\ 
         RCC & Train & $4$ & $587$\\ 
         RCC & Heldout & $3$ & $905$\\ 
         Ovarian & Train & $17$ & $1,995$\\ 
         Ovarian & Heldout & $4$ & $456$\\ 
         Prostate & Train & $1$ & $254$\\ 
         Prostate & Heldout & $2$ & $590$\\ 
         NASH & Train & $3$ & $644$\\ 
         NASH & Heldout & $2$ & $410$\\ 
         Lung & Train & $24$ & $3,859$\\ 
         Lung & Heldout & $19$ & $2,685$\\ 
         Esophageal & Heldout & $2$ & $893$\\ 
         Leukemia & Train & $1$ & $491$\\ 
         Lupus & Train & $3$ & $1,085$\\ 
         Bile Duct & Train & $3$ & $839$\\ 
         UC & Train & $2$ & $774$\\
         Other & Heldout & $4$ & $1,050$\\
    \thickhline
    \end{tabular}}
    \caption{Our proprietary Nuclei segmentation data characterization (includes H\&E and IHC).}
    \label{tab:nucseg}
\end{table}

\end{document}

%% file: preamble.tex
\usepackage[dvipsnames]{xcolor}
